\documentclass[%
reprint,
amsmath,amssymb,aps,
]{revtex4-1}
\usepackage{graphicx}% Include figure files
\usepackage{epsfig}
\usepackage{dcolumn}% Align table columns on decimal point
\usepackage{bm}% bold math

\begin{document}
\preprint{APS/123-QED}
% \title{Electron-Dominated Spontaneous Bifurcation of Harris Equilibrium}% Force line breaks with \\
%\thanks{Spontaneous Current Bifurcation of Harris Sheet Equilibrium}%
\title{Spontaneous Bifurcation of Single Peaked Current Sheets by Chaotic Electron
Scattering}
\author{Kuang-Wu Lee}
\email{lee@mps.mpg.de}

\author{J\"org B\"uchner}%
\affiliation{Max-Planck-Institut f\"ur Sonnensystemforschung,
37191 Katlenburg-Lindau, Germany}
\date{\today}% It is always \today, today,
\begin{abstract}
It is shown that single-peaked collisionless current sheets in a Harris-type
equilibrium spontaneously bifurcate as a result of chaotic scattering of
electrons at fluctuating magnetic fields near the center of the sheet,
as demonstrated by a 2D kinetic particle-in-cell simulation.
For this effect to be simulated explicit particle advancing is necessary,
since the details of the electron motion have to be resolved.
Unlike previous investigations of triggering bifurcated current sheet (BCS)
where initial perturbations or external pressure was applied the bifurcation
is spontaneous if thermal noise is taken into account. A spontaneous current
sheet bifurcation develops quicker than a tearing mode or other plasma instabilities.
It is shown that in the course of the current sheet bifurcation the Helmholtz free energy decreases
while the entropy increases, i.e. the new, bifurcated current sheet is in a more propable
state than the single-peaked one.

\begin{description}
\item[PACS numbers: 05.70.-a, 05.70.Ce, 52.35.Ra]
\end{description}
\end{abstract}
\pacs{Valid PACS appear here}% PACS, the Physics and Astronomy
\maketitle

%%%%%%%%%% introduce of current sheet equilibrium%%%%%%%%%%
The stability and possible unstable decay of current sheets play a central
role in astrophyics as well as in the laboratory, e.g. for magnetic energy
dissipation and reconnection \cite{Buechner:2006}.
For the investigations of current sheet stability
often single-peaked current sheets are used, as the one derived
by \citet{Harris:1962}.

%%%%%%%%%% instability of equilibria      %%%%%%%%%%
The free energy of equilibria can cause a number plasma instabilities which spontaneously
arise from thermal noise like the dissipative tearing mode instability in resistive \cite{Furth:1963}
and collisionless plasmas \cite{Coppi:1966}. In fusion plasmas spontaneous magnetic reconnection is
found recently in the poloidal current sheet, which is an initial equilibria in the Reverse Field Pinch
(RFP)\cite{Zuin:2009}. Since the single-peaked current sheet equilibria is unstable
and magnetic reconnection frequently takes place, the equilibria stability
is an important issue for current sheet evolution.

%%%%%%%%%% discovery of BCS in space%%%%%%%%%%
Magnetosphere is a natural plasma laboratory for the evolution of current sheet equilibrium.
Indeed, more recent detailed investigations of current sheets, has shown that current sheets
frequently are bifurcated (BCS) instead of single-peaked \citep{Sergeev:1993}
\citep{Hoshino:1996}\citep{Israelevich:2007a}. It was also found that
these BCS are electron dominated, e.g. by statistical analyses of
measurements onboard the CLUSTER spaccraft mission \citep{Israelevich:2007b}.

%%%%%%%%%% explain BCS as spontaneous process %%%%%%%%%%
BCS were also found in numerical simulations. In magnetic reconnection plan BCS was
interpreted, e.g., as a pair of slow mode shocks which develop in the
reconnection outflow region \citep{Shiota:2005}\citep{Thompson:2006}.
However, BCS were also observed in minimum plasma inflow conditions, for
which magnetic reconnection is not expected \citep{Tang:2006}.
In the current direction, BCS formation without plasma inflow is discovered
as a result of anomalous momentum transport due to pressure-gradient
driven lower hybrid drift instability \cite{Daughton:2004}.

\begin{figure*}
\epsfig{file = 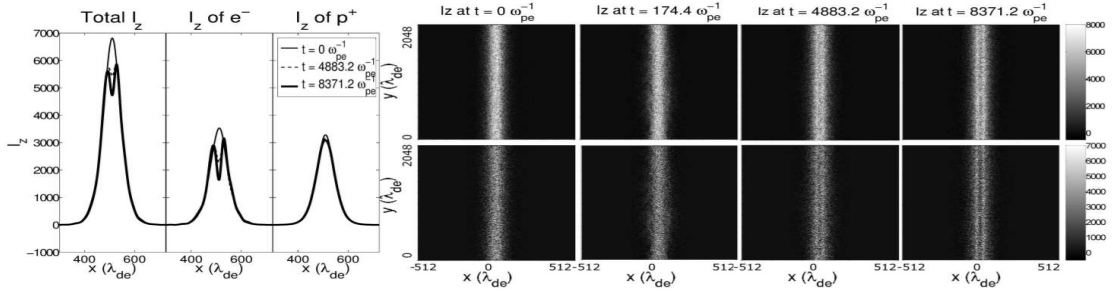, width=150mm, height=40mm}
\caption{The $y$-integrated current profiles (left panel)
and current density at four different simulation times
between $t=0$ and $t = 8371,2 \ \omega_{pe}^{-1}$. The
upper right panels show the total current and lower panels
electron current densities.
}
\label{fig1}
\end{figure*}

%%% previous numerical work on BCS, implicit scheme, initial perturbations %%%%%%%%%%
In the direction perpendicular to current drift, current sheet splitting similar to
BCS is observed after the saturation of tearing mode instability \citep{Lapenta:2005}.
Note that these authors used an implicit numerical scheme for advancing the particles
for their 2D particle-in-cell (PIC) code simulations.
In order to initialize the tearing mode instability quickly they also
imposed perturbations to trigger instability growth.

\citet{Schindler:2008} performed a one-dimensional particle-in-cell (1D PIC) simulation
with an initial boundary pressing. They concluded that a quasisteady boundary compression
forces a single-peaked current sheet evolves toward BCS, as an equilibrium relaxation
process but not due to plasma instability

%%%%%%%%%% questions to be answered and numerical approach %%%%%%%%%%
We now have found that single-peaked collisionless current
sheets might spontaneously bifurcate, without boundary compression or imposing tearing
mode perturbation initially, as long as the electron thermal fluctuations are considered
properly. We found that magnetic field fluctuations can initially start from thermal noise at
the center of current. The magnetic field fluctuations may lead to chaotic scattering
of the electrons out of the central (current peak) region of the sheet \cite{Buechner:1989},
reducing the electron current flow there and adding current flows away from the center of
the sheet as already discussed for laminar current sheets \cite{Zelenyi:2003}.

%%%%%%%%%% simulation background and setup %%%%%%%%%%
To prove this hypothesis quantitatively we carried out 2D electromagnetic
particle-in-cell code simulations to investigate the evolution of Harris
current sheet equilibrium. An explicit numerical scheme (XOOPIC) was implemented
since the details of the electron cyclotron trajectories have to be calculated
properly. The fastest processes up to electron plasma time scales
$\ \omega_{pe}^{-1} = \sqrt{\epsilon_o m_e / n e^2} $ is resolved.
The chosen current sheet has a width $2\lambda = 1.15 d_{i}$ which is slightly
larger than the ion inertial length $d_{i} = c / \ \omega_{pi}$ to cover the full
width of the ion dissipation region of the current sheet.
The equilibrium magnetic field of a Harris sheet equlibrium is
$B_{y}(x)=B_{0} \, tanh(x/\lambda)=\sqrt{4\mu k_{B} \ T \ N_{0}}
tanh(x/\lambda)$, where
$k_{B}$, $T=T_{i}=T_{e}$, $B_{0}$ and $N_{0}$ are the Boltzmann constant,
ion/electron temperatures, asymptotic magnetic field and number density at
the center of the sheet, respectively. The ratio of electron plasma frequency
to electron cyclotron frequency is $\ \omega_{pe}/\ \omega_{ce}=2.87$.
The ion to electron mass ratio used is $m_{i}/m_{e}=180$, which is
sufficient to separate the ion/electron motions. The grid size is of
the Debye length $dx=dy=\lambda_{De}=(\varepsilon_{0}T/Ne^{2})^{1/2}$, which
has sufficient spatial resolution of the detailed electron motion.
The simulation domain in the direction perpendicular to the current sheet
was $L_{x}=13.3d_{i}$ and in the current sheet direction $L_{y}=26.6d_{i}$.
This choice of $L_{y}$ allows a free development of the fastest growing
tearing-type and other eigenmodes of the sheet (see, e.g., Eq.(23) in
\citep{Brittnacher:1995}). The boundary conditions for the particles and
fields were chosen to be periodic in the $y$ direction and conducting walls
with particles reflection in the $x$ direction.

%%%%%%%%%% advantage of explicit scheme and no initial perturbation %%%%%%%%%%
The time step used was $dt=0.0872 \ \omega_{pe}^{-1}=0.03\ \omega_{ce}^{-1}$ to
fulfill Courant condition and to resolve the detailed electron oscillation and
cyclotron motions.
Note that this time step is much smaller than the one used in implicit
numerical schemes, where, e.g., in \citep{Lapenta:2005}
$dt=0.1\ \omega_{pi}^{-1}\approx 1.34\ \omega_{pe}^{-1}\approx 0.5\ \omega_{ce}^{-1}$
which does not track down the electron oscillation and cyclotron motion. The later
is significant for magnetic scattering, which we will discuss soon.
No initial perturbation was imposed as in \citep{Lapenta:2005}.
Also, no boundary compression was imposed as in the 1D PIC simulation
of BCS formation (see Fig.1 in \citep{Schindler:2008}).

%%%%%%%%%% simulated BCS and its characteristics %%%%%%%%%%
The left panels of Fig.\ref{fig1} show the $y$ integrated total current density
$I_z(x)=I_{z,e}(x)+I_{z,i}(x)$ and the individual contributions of electron $I_{z,e}$ and ion
$I_{z,i}$: for the initial, single-peaked Harris-equilibrium currents by a thin solid line, for
$t = 4883.2 \ \omega_{pe}^{-1}$ by a dashed line and for $t = 8371.2 \ \ \omega_{pe}^{-1}$
by a thick solid line.
At the late stage $t=8371.2 \ \omega_{pe}^{-1}\approx16.212\ \omega_{ci}^{-1}$
two current peaks are formed which are well separated from each other.
As one can see already in the middle panel of the left Fig.\ref{fig1}
the reduction of $I_z(x)$ near the center of the current sheet is mainly
due the spatial redistribution of the electron current. Since the ion
current is not changing much before $t=8371.2 \ \omega_{pe}^{-1}$ the right
panels of Fig.\ref{fig1} depict the evolution of the total current density
distribution in the $x, y$ plane with time in the upper row and the
responsible for the current redistribution electron part in
the lower panels.
While initially both the total and the electron currents are concentrated
near the center of the sheet (first column in the right part of
Fig.\ref{fig1}), already at
$t=4883.2\ \omega_{pe}^{-1} = 9.5 \ \omega_{ci}^{-1} = 1.5 \tau_{ci}$
a clear dip of the current density has developed
around the center of the sheet. Here $\tau_{ci} = 2 \pi
\ \omega_{ci}^{-1} $ denotes a cyclotron period. Hence, the bifurcation
takes place at the time scale of an ion cyclotron period.
This result indicates that the ion-to-electron mass ratio $m_{i}/m_{e}=180$
used in the simulation the electron and ion dynamics are sufficiently
well separated.

%%%%%%%%%% current profile bifurcates earlier than previous works %%%%%%%%%%
Note further that the bifurcation due to boundary pressing takes place
only after $t\approx \ 400 \ \omega_{ci}^{-1}$ \citep{Schindler:2008}.
Their mass ratio is $m_{i}/m_{e}=25$ and the bifurcation time in electron
plasma period is $400 \ \omega_{ci}^{-1}\approx \ 4.4 \times 10^{4}\ \omega_{pe}^{-1}$.

\begin{figure}
\epsfig{file = 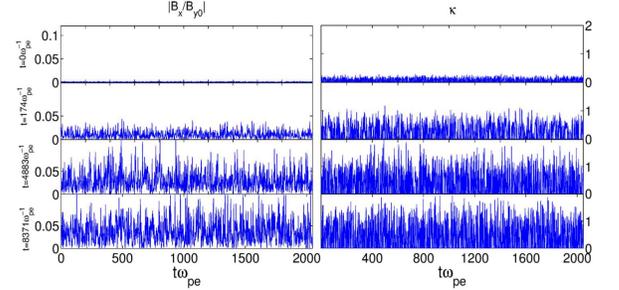, width=80mm , height=40mm}
\caption{Absolute values of the fluctuating $\vert B_x \vert$ magnetic fields
at the center of the current sheet, normalized to the asymptotic magnetic
field of the Harris equilibrium (left panel) and the thermal electron
$\kappa (y)$ values at center of the current sheet (right panel) for
the same four simulation times as in Figure 1, right panels.}

\label{Bxkappaprofile}
\end{figure}

%%%%%%%%%% BCS resulted from chaotic scattering %%%%%%%%%%
In order to understand this phenomenon one has to realize that plasmas
are coarse-grained by their particles whose thermal motion lets currents and
magnetic fields fluctuate at small scales.
Hence, while the self-consistent magnetic field $B_y$ of a Harris-equilibrium
vanishes at the center of the sheet, fluctuating magnetic fields remain.

The left panel of Fig.\ref{Bxkappaprofile} shows $\vert B_x(x=0) \vert$,
the absolute value of the normal field $B_x$ component of the magnetic field
at the center of the sheet, divided by the asymptotic Harris sheet
field $B_{yo} = B_0$ for the same simulation times as chosen for
the right panels in Fig. 1.
As one can see, $\vert B_x(x=0) \vert$ is finite most of the time unless
the particles' thermal velocity is not taken into account as in the the
uppermost column ($t=0$). It is well known that particles transiting the center of
current sheets can be chaotically scattered of the curvature of the
magnetic field is comparable to the Larmor radii. According to
\citet{Buechner:1989} the scattering is strongest, when the parameter
$\kappa = \sqrt{R_{min}\rho_{max}}$ is unity, where $R_{min}$ and $rho_{max}$
are the minimum curvature of the magnetic field and and the maximum particle
gyroradius at the current sheet center. The right panel of
Fig.\ref{Bxkappaprofile} depicts the $\kappa$ values for thermal
electrons along the y direction for the same moments of time for which
the $B_x(y)$ fields are shown in the left panel of this Figure and in the
right panel of Fig.1.
As one can see in the right panel of Fig.\ref{Bxkappaprofile} before
$t  = 4883 \ \omega_{pe}^{-1}$  $\kappa$ reaches unity only at a few
positions, i.e. only at a few places the electrons are strongly scattered.
After $t  = 4883 \ \omega_{pe}^{-1}$ electrons are strongly scattered
chaotically at many positions.

The consequences of this strong scattering can be described best by
means of the action integral of the fast motion motion
$I = \oint v_z dz$ \cite{Sonnerup:1971}. For $\kappa < 1$ this
action integral can be expressed in its normalized form as
$I' = \left( { \frac {\kappa^2 y'} {2 k^2 -1} } \right)
\ f_{A/B}(k)$, where $k(y')$ is a function of $y'$, the appropriately
normalized $y$ coordinate of the slowly moving guiding center and
$f_{A/B}(k)$ are two different functions of complete elliptic integrals
for $k < 1$ and $k > 1$, respectively \cite{Buechner:1989}.
For $\kappa < 1$ the action integrals $I'$ are adiabatically conserved,
i.e. integrals of motion, as long as they stay away from the separatrix
which is reached when $k \to 1$. For $k < 1$ such quasi-adiabatic orbits
cross the center of the current sheets while for $k > 1$ they will gyrate
at some distance from the sheet center. Particles with $1 < I' <
I'_{max} = 1.16$ are trapped on crossing the current sheet orbits $k < 1$.
But in thin current sheets most particles will have $I' < 1$, which
enables them to eventually reach a separatrix in the velocity space, i.e.
change from the region $k < 1$ to $k > 1$, at $k=1$ changing from meandering
across the sheet, drifting in the direction of the original sheet
current to a gyration away from the current sheet center where they
drift in the opposite direction, causing diamagnetic currents away
from the sheet center \cite{Buechner:1989}.
While for $\kappa \ll 1$ particles the value of the quasi-integral of
motion $I'$ stays practically unchanged during the sepratrix encounter,
for $\kappa \to l$ it can be essentially changed. If the obtained value
of $I'$ is very small only a small amount of kinetic energy is left in
the perpendicular to the magnetic field velocity direction while most
energy is in the slow drift motion.
These particles contribute significantly to the built up of diamagnetic
currents away from the sheet center. For small $I'$ the asymptotic
expressions for the elliptic integrals reveal
$I' \approx 3 / 16 \pi k_{tp} \kappa^6$, where $k_{tp}$ correspond to the
turning point of the drift motion in $y'$ ($v_{y'} = 0$).
From the condition for turning the drift in the $y$ direction
one obtains the position of the turning point by solving the
equation $\kappa^2 y'_{tp} = 2 k_{tp}^2 -1$. Solving this equation
for $k_{tp}$s and calculating the mean $z$-position for the
turning point location one obtains a the distance $\Delta z  \sqrt{\lambda
\rho_{the}} (16 \pi / 3)^2 I'_{the}$ from the current sheet at which
most of the particles drift in the dia-magnetic current direction.
Here $I'_{the}$ means the quasi-adiabatic invariant for typical thermal
electrons, the bulk drift velocity of the electrons is much smaller
and can be neglected. With the time being $\Delta z $ will become the
position of the maximum of the dia-magnetic flow of the electrons
where the initial current profile is modified most by electron flows
while the electron flow at the center of the current sheet is reduced.
For the parameters used in the simulation one obtains $\Delta z \cong 27
\lambda_{De}$. This theoretically predicted distance corresponds to the
one found in simulation.

\begin{figure}
\epsfig{file = 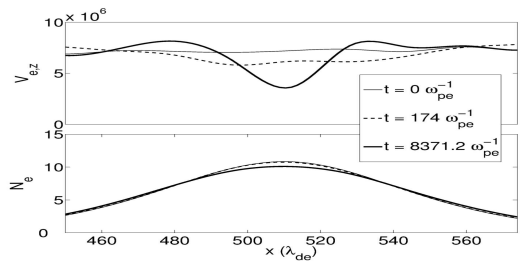, width=70mm, height=35mm}
\caption{Profiles of electron drift velocity $V_{e,z}$
(upper panel) and number density $N_{e}$ (lower panel), both
integrated along the $y$ axis.}
\label{electrondriftdensity}
\end{figure}

\begin{figure}
\epsfig{file = 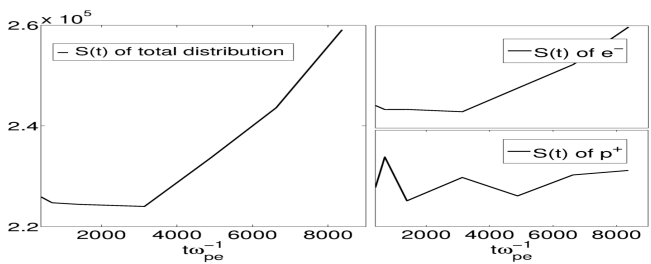, width=90mm, height=35mm}
\caption{Evolution of the total relative entropy (left panel)
and of the electron (upper right panel) and ion entropies
(lower right panel) separately.}
\label{entropy}
\end{figure}

The $Y$ integrated electron drift $V_{e,z}$ and density $N_{e}$
profiles are show in Fig.\ref{electrondriftdensity} for three moments of time.
Note that such current profile was obtained by \citet{Hoshino:1996}
from a statistical analysis of many current sheet encounters in the Earth's
magnetotail.
The electron number density does not change much in the course of the
bifurcation, which happens, instead, in the electron drift velocity space.
This is consistent with the finding that the current
depletion at the center is caused mainly by the redistribution of the electrons
by pitch-angle scattering in the velocity space when crossing the sepratrix
between meandering ($k < 1$) and gyrating away from the sheet center ($k > 1$).
Due to the mass ratio only the electrons undergo strong chaotic scattering
thin Harris-type current sheets.
%%%%%%%%%% BCS are also equilibrium %%%%%%%%%%
The new, bifurcated current sheet is again in equilibrium. In fact there is
an infinite number of possible current sheet equilibria. A number of
analytical \cite{Neukirch:2009} and non-analytical equilibrium solutions have
been found which are more realistic for space current sheets than Harris
sheets \cite{Cowley:1978}. For a survey see, e.g., \cite{Schindler:2002}.
A current sheet bifurcated out of a single peaked Harris current sheet
by the chaotic electron scattering is close to the equilibrium found
by \citet{Lapenta:2005}, their case (b).
%%%%%%%%%% Is the BCS in simulation a more stable state? %%%%%%%%%%
Note that the BCS, naturally obtained via chaotic electron scattering, is
more probable than the single peaked Harris sheet. This can be demonstrated
by calculating the entropy of the system. For a plasma with a continuous
particle distribution, it is appropriate to consider the relative
Kullback-Leibler entropy which is always positive \citep{Kullback:1951}.
With respect to a reference distribution $q(v)$ at $t = 0$ it can be
written as
\begin{eqnarray}
\nonumber
S_{KL}(t) = \int_{-\infty}^{\infty}dv f(v,t) ln (\frac{f(v,t)}{q(v)|_{t=0}})
\end{eqnarray}
The relative entropy as a sum of
the electron and ion contributions grows increases in the course of current
sheet bifurcation. Looking at the electrons and ions separately one can
see that the entropy of electrons is, indeed, steadily increasing
after about $t=3800 \ \omega_{pe}^{-1}$ while the ion entropy oscillates
(right columns of Fig.\ref{entropy}).
%%%%%%%%%% Helmholtz free energy calculation indicates BCS is more stable %%%%%%%%%%
The stability of a steady state equilibrium can be analyzed by calculating the
Helmholtz free energy $F = U -T \ S$ \citep{Kan:1972}. Here the $U$ is the internal
energy, a sum of the particles kinetic energy and the field energy, $T$ is
the temperature energy and $S$ is the entropy. Less free energy $F$ corresponds
to a more stable equilibrium.
In the course of the bifurcation, the internal energy U is conserved
while T and S are increasing. Hence the bifurcated current sheet
contains less free energy. This is why it is more favored in nature
than single-peaked equilibria like the Harris current sheet and it is
more stable.

%%%%%%%%%% summary %%%%%%%%%%
2D PIC simulations confirmed that an initially single
peaked Harris equilibrium current sheet does spontaneously
and quickly bifurcates a natural consequence of chaotic
electron scattering due to their fluctuations at the center of the
sheet. The bifurcation is faster than spontaneous plasma
instabilities as of the tearing mode. The single peaked sheet
bifurcates without initial perturbations or imposed external
pressure. The bifurcated state is more favorable because
electrons gaining after chaotic scattering at the sheet center
toward small values of the quasi-adiabatic invariant
of motion $I'$ spends longer time in the cucumber
phase of their motion away from the sheet center rather
than meandering at the sheet center.
Therefore it can be simulated only by
numerical schemes that explicitly resolve the details of
the electron motion.
The spontaneous bifurcation of single peaked current sheets explains
the frequent observation of bifurcated current sheets in quite
situations without plasma inflows and reconnection as they are
more stable than single peaked sheets.
Their influence on plasma instabilities is stabilizing, delaying,
lowering the growth rate, e.g., of the tearing mode instability.
%%%%%%%%%% acknowledgement %%%%%%%%%%
\begin{acknowledgments}
The authors are grateful to the Max-Planck Society for funding "Turbulent
transport and ion heating, reconnection and electron
acceleration in solar and fusion plasmas" Project No. MIF-IF-A-AERO8047.
\end{acknowledgments}

%\bibliography{BifurcationJB4}% Produces the bibliography via BibTeX.

\end{document}